\documentclass{aa}
\usepackage{epsfig}

\def\iue{\mbox{\it IUE}} 
\def\ines{\mbox{\it INES}} 
 
\def\newsips{\mbox{\it NEWSIPS}} 
\def\iuesips{\mbox{\it IUESIPS}}

\def\back2{\mbox{$B(x,\lambda)$}} 
 
\def\sig2{\mbox{$\sigma(x,\lambda)$}}

\begin{document} 

\thesaurus{23(03.13.2;03.20.1;13.21.2)}

\title{The INES system. II:
Ripple Correction and Absolute Calibration for IUE High Resolution Spectra}

\author{
A. Cassatella\inst{1,2}\fnmsep $^*$
\and A. Altamore\inst{2}
\and R. Gonz\'alez-Riestra\inst{3}\fnmsep\thanks{{\it Previously:} ESA--IUE Observatory}
\and J.D. Ponz \inst{4}
\and J. Barbero \inst{5}
\and A. Talavera \inst{3}\fnmsep $^*$
\and W. Wamsteker \inst{6}\fnmsep\thanks{Affiliated to the Astrophysics Division, 
SSD, ESTEC}
}

\offprints{A. Cassatella}
\mail{cassatella@fis.uniroma3.it}

\institute
{
CNR, Istituto di Astrofisica Spaziale, Via del Fosso del Cavaliere
100, I~00133 Roma, Italia
\and Dipartimento di Fisica E. Amaldi, Universit\'a Roma  Tre, Via della
Vasca Navale 84, I~00146 Roma, Italia
\and Laboratorio de Astrof\'{\i}sica Espacial y F\'{\i}sica Fundamental, 
P.O. Box 50727, E~28080 Madrid, Spain
\and European Space Agency, Villafranca Satellite Tracking  Station, P.O. Box 50727, E~28080 Madrid, Spain
\and Departamento de F\'{\i}sica Aplicada, Universidad de Almer\'{\i}a, 
Carretera de Sacramento s/n, E~04120 Almer\'{\i}a, Spain 
\and ESA-IUE Observatory, VILSPA, P.O. Box 50727, E~28080 Madrid, Spain
}

\date{Received / Accepted}

\authorrunning{A. Cassatella et al.}

\titlerunning{Ripple correction of IUE High resolution spectra}

\maketitle

\begin{abstract}

In this paper we document the results of the study which led to
the ripple correction and absolute calibration algorithms
applied to the high resolution spectra processed with the
\newsips\ software for  the Final Archive of the \iue\ Project. 
In this  analysis,   based on a very large number of spectra, we
find that both K and the $\alpha$ parameters (not only the
former as previously believed) vary with order number. This
fact, together with the finding that the central peaks of the
blaze function vary also as a function of the THDA temperature
(for the SWP camera) and of the date of observations (for the
LWP and LWR cameras), makes the ripple correction algorithm more
complex than previously considered but, at the same time,
considerably more reliable. As for the high resolution absolute
calibration, the method followed is similar to the one
implemented in \iuesips. The internal accuracy of the high
resolution calibration is about 4\%. We note that the ripple
correction and absolute calibration algorithms here described
apply also to \iue\ data processed and distributed with the
\ines\ system.


\end{abstract}

\section{Introduction}
\label{sec:intro}

In this paper we address two fundamental aspects which
influence, to different extent, the photometric quality of \iue\
high resolution spectra, i.e. the correction for the echelle
blaze function (ripple correction), and the absolute
calibration. The results of this study, which are based on about
one thousand  spectra, mainly of \iue\ standard stars, have been
implemented in  the \iue\ Newly Extracted Spectra system
(\ines), a data archive and distribution facility developed by
the \iue\ ESA project (see Wamsteker et al. 1999), as well as in
the \newsips\ data reduction package (Nichols and Linsky 1996, 
Garhart et al. 1997).  Full details on the \ines\ system are
provided by Rodr\'{\i}guez-Pascual et al. (1999, Paper I) for
low resolution data, and by Gonz\'alez-Riestra et al. (1999a,
Paper III) for high resolution data.

\section{The NEWSIPS Ripple Correction}
\label{sec:ripple}

The correction of \iue\ high resolution spectra for the echelle
grating efficiency has for long been considered a critical area 
in the context of \iue\ data reduction procedures. The \ines\ 
system includes an upgraded  ripple correction algorithm which
is presented and discussed in the following.

\subsection{The method}

Let us indicate with R$_m(\lambda$) the grating efficiency
(blaze function) of the high resolution spectrographs, for a
given order m and  wavelength $\lambda$. As shown by Ahmad
(1981)  and Ake (1981), this function is adequately represented
by:

\begin{equation}
\rm R_{\rm m}(\lambda)=\sin^2{\rm x/x}^2
\label{eq:rip}
\end{equation}

\noindent 
where

\begin{equation}
\rm x=\pi~\rm m~\alpha~(1-\lambda_{\rm c}(m)/\lambda)
\end{equation}

\noindent 
being $\lambda_c$(m) the wavelength of the central maximum of
the blaze function for order m, and $\alpha$ a parameter which
can be shown to be, in the first approximation,  inversely
proportional to the  Half Width at Half Maximum (HWHM) of the
order considered:

\begin{equation}
\rm HWHM=1.395~ \lambda_c(m)/(\pi~m~\alpha) 
\end{equation}

\noindent
The quantity

\begin{equation}
\rm K = m \lambda_c(m) 
\end{equation}

\noindent 
is called the 'ripple constant', but, as discussed in the
following, it turns out to be a function of order number for the
\iue\  setup.

Let us indicate with f($\lambda$) the observed net spectrum
normalized  to the exposure time. The shape of the blaze
function  R$_{\rm m}(\lambda$) can be determined  from the
observations,  knowing the inverse sensitivity curve of the
camera (S$^{-1}$($\lambda$)C($\lambda$), see Section
\ref{sec:abscal}), and the absolute flux of the target
F($\lambda$):

\begin{equation}
\rm R_{\rm m}(\lambda)=\rm f(\lambda)~\rm C(\lambda)~\rm S^{\rm -1}(\lambda)/\rm F(\lambda)
\label{eq:erre}
\end{equation}

\noindent
This equation implies that the blaze parameters cannot be
derived from  f($\lambda$) directly, because it is affected by
the distortions introduced  by the multiplying factor
C($\lambda$)S$^{-1}$($\lambda$)/F($\lambda$), which is not only 
wavelength--dependent, but  differs also from target to target.
Not having taken this fact into account is one of the major
causes for the inaccuracy of the ripple correction in previous
releases of the \iue\ data reduction packages and, in
particular, for the suspected dependence of the ripple
parameters on the spectral type of the source considered.

Our approach to derive the ripple parameters is  equivalent to
using Eq.  \ref{eq:erre}, but is of much easier  implementation.
It consists in normalizing the observed spectrum to its
continuum. The continuum was determined, for each spectral
order,  in correspondence to the expected peaks of the blaze
function (Grady and Garhart 1989), where $R_{\rm m}(\lambda)
\simeq  1$, and then  interpolated over wavelength. 

\begin{figure}
\psfig{file=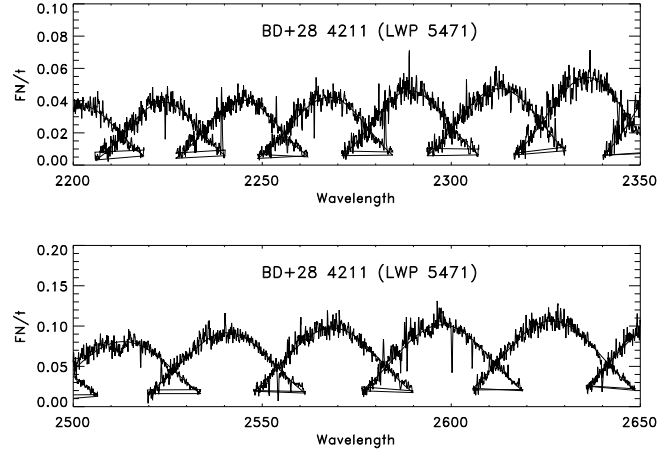,width=9cm}
\caption{Example of fitting Eq.~\ref{eq:rip} to a portion of a
LWP spectrum of BD+28~4211. }
\label{fig:example}
\end{figure}

Let us indicate with Y($\lambda$) the net spectrum extracted
from order m after normalization to the continuum, and with 
T(m,$\lambda$) its analytical representation. We can write:

\begin{equation}
\rm T(m,\lambda)=\rm A_{\rm m}~R_{\rm m}(\lambda)
\end{equation}

\noindent
where A$_{\rm m}$ is an adjustable parameter representing the
net peak intensity of the blaze function (at
$\lambda=\lambda_{\rm c}$). Because of normalization to the
continuum, A$_{\rm m} \simeq 1$.  The unknown blaze parameters
$\alpha$, $\lambda_{\rm c}$(m) and A$_{\rm m}$ were determined
iteratively, for each order, via a non-linear least-squares
fitting algorithm, i.e. by minimizing $\sum{\rm (T_i-Y_i)^2}$. 
All the data points were given the same weight except those
affected by saturation, ITF extrapolation, reseau marks or
particle events, which were discarded.  Spectral absorption
features were also discarded through an automatic rejection
procedure. An example of the fitting for the  LWP05471 spectrum
of the standard star BD+28~4211 is shown in
Fig.~\ref{fig:example}.

We stress that the ripple parameters should be derived, as done
here,  in vacuum wavelengths and in the velocity scale of the
\iue\ spacecraft.

\subsection{The data base}

For this study we have selected 516 good quality high resolution
spectra of the \iue\ calibration standards obtained between 1978
and 1995, and namely: 182 spectra for the SWP camera, 296 for
the LWP and 138 for the LWR. This represents the near totality
of data of this kind available. As input data we used the MHXI
net spectra as extracted with \newsips. Among the most
frequently  observed standards we quote  BD+28 4211, BD+75 325,
HD60753, HD3360, G191 B2B, NGC246, HD120315 and   HD93521. In
addition, to test specific aspects of the ripple  correction 
algorithm, it was necessary to introduce  about 200 net spectra
(also extracted with \newsips) of targets having heterogeneous
spectral types. The large majority of the spectra were obtained
through the  large entrance apertures of  the \iue\
spectrographs. A few small aperture spectra were also
considered, whenever available, to cross-check the applicability
of the large aperture  ripple correction algorithm to that case.
For the long wavelength cameras, several overexposed spectra
were also included to refine the ripple parameters at the
shortest wavelengths, where the cameras sensitivity is low.

\subsubsection{SWP}

\paragraph {Dependence of K on order number:}
We find that the central positions  of spectral orders
$\lambda_c$(m) are not stable for SWP spectra, as commonly
believed in the past but, for a given order, vary strongly as a 
function of the Camera Head Temperature (THDA). Given this, a
convenient way to proceed is to  determine first 
$\lambda_c$(m)   from data obtained within a restricted range of
THDA values.  We used 52 spectra of the  stars BD+28$^o$4211 
(13 spectra),  BD+75$^o$325 (20 spectra), CD-38$^o$10980 (5
spectra), HD60753 (12 spectra) and NGC246 (2 spectra) obtained
the range $8.5  \le THDA \le 11.5^o$C.  Being the mean value of
THDA for this sample,  10.04$\pm$0.70 $^{\rm o}$C,   very close
to the average operating conditions for the SWP camera during
years 1978 to 1991, the corresponding central positions 
$\lambda_c$(m)   can be used to obtain a mean curve  
K=m~$\lambda_{ref}$, where  $\lambda_{ref}$ =
$\lambda_c$(m,THDA=10.04) are the central wavelengths   to be
taken as a reference for that temperature. A linear regression
to the mean values for the  four stars provides:

\begin{equation}
{\rm K=m}\lambda_{\rm ref}=137508.316+2.44761341{\rm m}
\label{eq:swpkap}
\end{equation}

\noindent
The associated uncertainty on the central wavelengths is 0.24
\AA\ at 1400 \AA\ and 0.31 \AA\ at 1850~\AA. The
ripple-corrected fluxes in the overlap region between adjacent
orders are accurate to within 3\%\ above 1700 \AA\ and 6\%\
around 1200 \AA. However, the flux mismatch in the overlap
region between adjacent orders caused by a systematic error on
the central wavelengths is twice the quoted errors, because if
the flux at order m is overestimated, the flux in the overlap
region of the adjacent order m-1 will be underestimated and
viceversa. 

\begin{figure}
\psfig{file=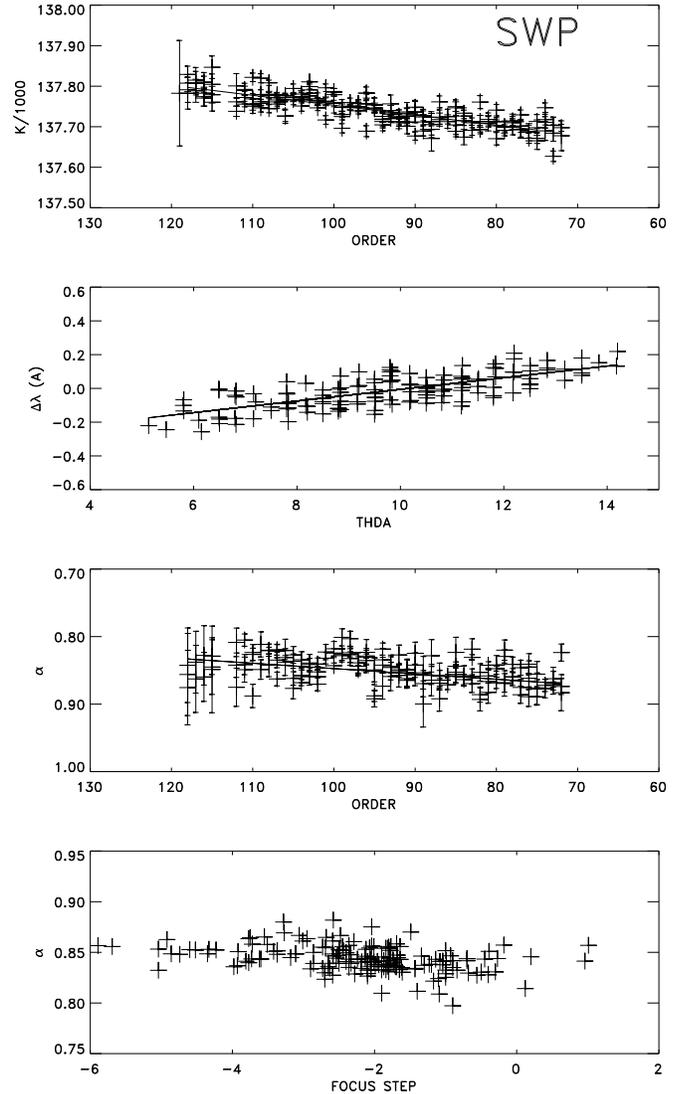,width=9cm}
\caption{Ripple parameters for the SWP camera. Form top to
bottom: Variation of K with order number, dependence of
$\Delta\lambda$ on THDA, and dependence of $\alpha$ on order
number and on   focus step.} 
\label{fig:swprip} 
\end{figure}

\paragraph {Dependence of $\lambda_{\rm c}$ on THDA:}

To extend the validity of Eq. \ref {eq:swpkap} to any regime of
THDA, we have used the whole set of input spectra and computed
for each spectrum the wavelength difference between the observed
central wavelengths and those obtained from Eq.
\ref{eq:swpkap}.  A linear fit to the data over the range of
orders 110-86, where the blaze function is narrower, and then
better sampled all over its shape, provides:

\begin{equation}
\lambda_{\rm c}-\lambda_{\rm ref}=-0.335775+0.0321729~{\rm THDA}
\label{eq:swpdel}
\end{equation}

\noindent
with a standard deviation of 0.069 \AA. This equation quantifies
the previously  quoted dependence of the central positions on
THDA.   We have also looked for a possible dependence of the
central  wavelengths on the epoch of the observation, on the
spectral type of  the target and on the exposure level, but no
correlation was found. \\

\paragraph{Dependence of $\alpha$ on order number and focussing conditions:}
We find that the ripple parameter $\alpha$ varies linearly as a
function  of the order number. A linear fit to the data
provides:

\begin{equation}
\alpha=0.92628-7.890132~10^4{\rm m}
\label{eq:swpalp}
\end{equation}

\noindent
with a standard deviation of 0.022. The corresponding
uncertainty on fluxes in the regions midway between adjacent
orders is 3.6\%, irrespective of the order considered. Note that
the errors in $\alpha$ do not produce any flux discontinuity in
the overlap region between adjacent orders.

No dependence of $\alpha$ on the date of observation or THDA 
was detected. On the contrary, there is a marginal indication 
of a slight  decrease of $\alpha$ with increasing focus STEP
parameter. This effect, if real, would imply that the width of
the spectral orders becomes larger as STEP departs from the
optimum focussing conditions. Since most \iue\ data are taken at
optimum focussing conditions and the effect is any case
marginal, it has not been implemented in the ripple correction
algorithm.  In Fig. \ref{fig:swprip}  we show the ripple
constant K as a function of  order number, the wavelength shift
$\Delta\lambda$ = $\lambda_{\rm c} - \lambda_{\rm ref}$ as a
function of THDA, and the $\alpha$ parameter as a function of
order number and focus STEP.

\begin{figure}
\psfig{file=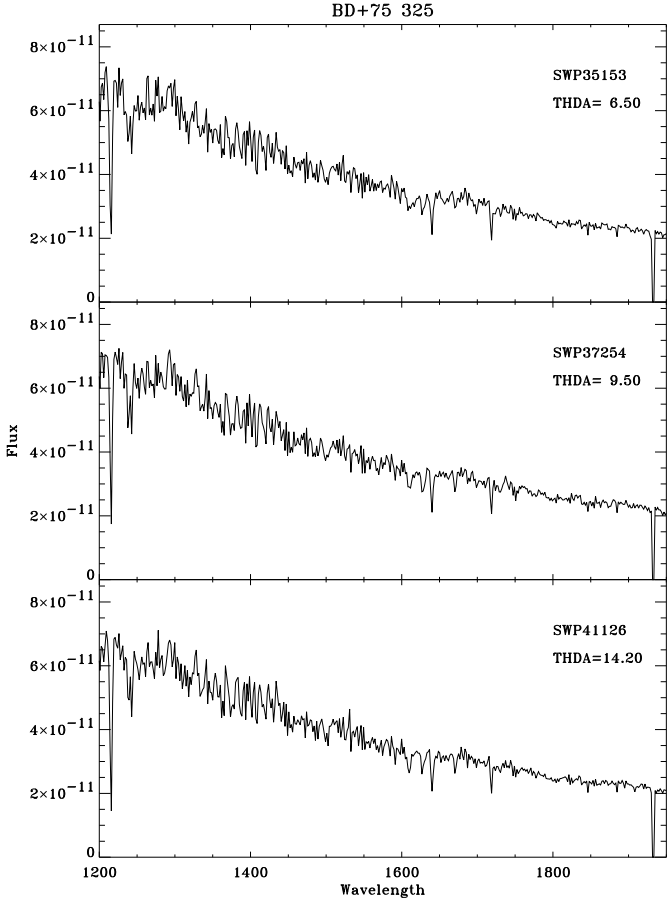,width=9cm}
\caption{Energy distribution of BD+75 325 obtained from three
SWP ripple corrected high resolution rebinned spectra obtained
under very different THDA conditions. Fluxes are in units of erg
cm$^{-2}$ s$^{-1}$ \AA$^{-1}$. }
\label{fig:swpridem}
\end{figure}

Given the strong dependence of the central wavelengths of
spectral orders on  the THDA temperature, we have verified  that
the SWP ripple correction algorithm works well even when the
THDA values  depart considerably from the average operating
conditions.  As an example, we show in Fig. \ref{fig:swpridem},
three spectra of BD+75 325 obtained with very different THDA
values (6.5, 9.5 and 14.2): no periodic flux fluctuations
(typical of bad ripple correction) are seen in the corrected
spectra.

\subsubsection{LWP}
 
\paragraph{Dependence of K on order number:}
Contrary to the case of the SWP camera, the central wavelengths
of LWP spectral orders $\lambda_c$(m) are not sensitive to the
camera head temperature THDA,  but depend strongly on the date
of the observations.

It is then convenient to study first the dependence of K on
order number  for spectra obtained in a sufficiently short  time
interval. We have selected the period from year 1988 to 1992
because many (86) good quality  spectra  of the \iue\
calibration standards were obtained.  The mean observing date of
this restricted sample was T = 1991.2 (fractional year), and the
mean and rms values of THDA and focus STEP were  10.18$\pm$1.81
$^o$C and -2.84$\pm$0.85, respectively.   Since the central
positions were sufficiently stable in the quoted period of time,
the values of K=m$\lambda_c$(m,T=1991.2) obtained from the
individual spectra were averaged together. A linear regression
to these data provides:

\begin{equation}
{\rm K}={\rm m}\lambda_{\rm ref}(m)=230868.17770+3.86260914{\rm m}
\label{eq:lwpkap}
\end{equation}

\noindent
where $\lambda_{ref}$=$\lambda_c$(m,T=1991.2) are the reference
central wavelengths for time T=1991.2.  The standard deviation
of the fit is 30.1, corresponding to an uncertainty on the
central wavelengths of 0.30 \AA\ at 2300 \AA. The ripple
corrected fluxes in a point midway in    the overlap region
between adjacent orders are accurate to within 3.9\%\ at 2200
\AA, and 2.3\%\ at 2800 \AA.

\begin{figure}
\psfig{file=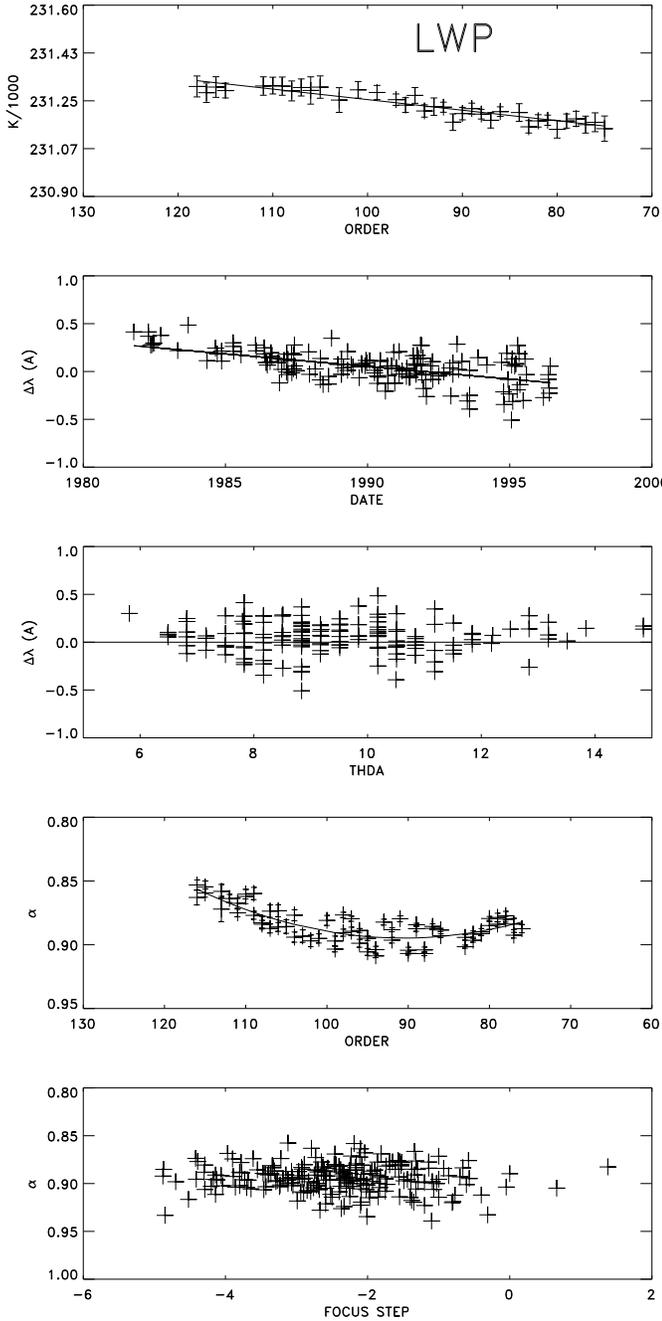,width=9cm}
\caption{Ripple parameters for the LWP camera. Form top to
bottom: the K parameter as a function of order number; the
wavelength shift $\Delta\lambda$ = $\lambda_{\rm c}
-\lambda_{\rm ref}$ as a function of observing date and THDA;
the $\alpha$ parameter as a function of order number and focus
STEP.  Note the lack of correlation between $\Delta\lambda$ and
THDA, and between $\alpha$ and focus STEP.}
\label{fig:lwprip}
\end{figure}

\paragraph{Dependence of $\lambda_{\rm c}$ on time:}
In this section we show  that the central wavelengths
$\lambda_{\rm c}$ of a given order vary with time, being their
deviations with respect to the mean values in 1991.2 (see Eq.
\ref{eq:lwpkap}) a linear function of observing time. To define
this dependence, we have selected 142 spectra obtained between
1981 and 1995, and  derived, for each order, the wavelength
difference with respect to the reference values for 1991.2.
Since the shape of the K(m) function is linear at any date and
with a  very similar slope,  the wavelength differences of
individual spectra  could be averaged together over the orders 
110 to 80, providing a mean wavelength shift  $\Delta\lambda$ =
$\lambda_{\rm c}$ -$\lambda_{\rm ref}$,  which is  plotted as  a
function of time  in Fig. \ref{fig:lwprip}.   A linear fit to
these data provides:

\begin{equation}
\lambda_{\rm c}-\lambda_{\rm ref}=52.570796-0.0263910~{\rm T}
\label{eq:lwpdel} 
\end{equation}

\noindent
with a standard deviation of 0.10 \AA. \\

\paragraph{Dependence of $\alpha$ on order number and focussing conditions:}
We find that the ripple parameter $\alpha$ shows a strong
dependence on the  order number. A linear fit to the data
provides:

\begin{equation}
\alpha=0.406835+0.01077191{\rm m}-5.945406~10^{-5}{\rm m}^2
\label{eq:lwpalp}
\end{equation}

\noindent
and a standard deviation of 0.00867. The corresponding
uncertainty on fluxes in the overlap region midway between
adjacent orders is 1.4\%.

We have also investigated to which extent the central
wavelengths depend on the camera head temperature THDA. To this
purpose we have selected 194 spectra of calibration standards
and computed the wavelength difference
$\Delta\lambda$=$\lambda_c-\lambda_{ref}$ averaged over the
order range 110-80. The results indicate that there is no
correlation between $\Delta\lambda$ and THDA, unlike the case of
the SWP camera, as shown in Fig. \ref{fig:lwprip}. 

Similarly, no correlation has been found between the $\alpha$ 
parameter and the focussing conditions, nor between the blaze
parameters and  the exposure time or the energy distribution of
the star.  The main results for the LWP camera are provided in
Fig. \ref{fig:lwprip},  which shows the K parameter as a
function of  order number, the wavelength shift
$\Delta\lambda$=$\lambda_{\rm c}-\lambda_{\rm ref}$  as a
function of observing date (fractional year) and THDA, and  the
$\alpha$ parameter as a function of order number and focus STEP.

\begin{figure}
\psfig{file=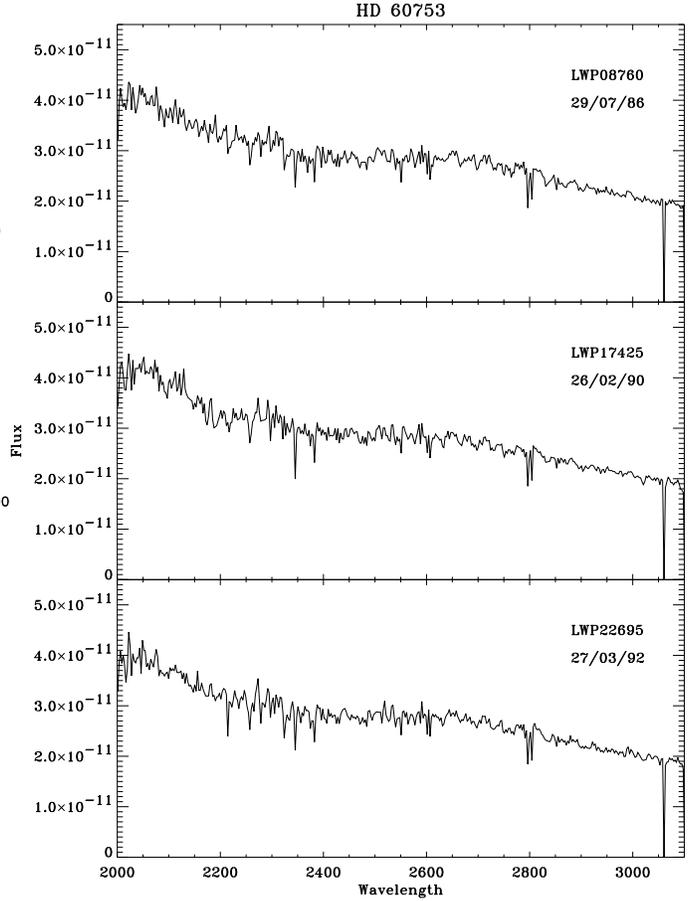,width=9cm}
\caption{Flux distribution of HD 3360 obtained from three LWP
ripple corrected and flux calibrated high resolution rebinned
spectra obtained at very different epochs. Fluxes are in the
same units as in Fig. \ref{fig:swpridem}.}
\label{fig:lwpridem}
\end{figure}

We find that, in spite of the strong dependence of  the central
positions of spectral orders on the time of observations (see
Eq. \ref{eq:lwpdel}), the quality of the ripple correction
remains good even for spectra obtained several years apart, as
shown in the example of Fig. \ref{fig:lwpridem}.

\subsubsection{LWR}

\paragraph{Dependence of K on order number:}
As for the LWP camera, the central wavelengths of LWR spectral
orders $\lambda_c$(m) depend strongly on the date of the
observations, but are insensitive to THDA variations. 

Following the same procedure used for LWP spectra,  we have in
first place determined a mean curve  K(m)= m$\lambda_{\rm c}(m)$
for a fixed epoch, to be taken as a reference.  A restricted
sample of 58  spectra of \iue\ calibration standards obtained
between 1982 and 1986  was used for this purpose.  The mean
observing date was T=1982.6 and the mean values of THDA  and
focus STEP were 13.40$\pm$1.70 $^o$C and -1.3$\pm$0.9,
respectively.  The central wavelengths of spectral  orders were
then averaged together to obtain a mean value of 
K=m$\lambda_c$(m,T=1982.6).  A fourth order polynomial fit to
the results provides:

\begin{equation}
{\rm K=A+B~m+C~m^2+D~m^3+E~m^4}
\label{eq:lwrkap}
\end{equation}

\noindent where \\

\noindent A=0.281749635~10$^6$ \\
B=-0.223565585~10$^4$ \\
C=0.365319482~10$^2$ \\
D=-0.262477775 \\
E=0.701464055~10$^{-3}$ \\

\noindent
The standard deviation of the fit (18.06) corresponds to an
uncertainty on the central wavelengths of 0.18 \AA\ at order
100. 

We find that the central wavelengths of LWR spectral orders
$\lambda_c$(m) do not show any dependence on the camera head
temperature THDA, as  for the LWP camera. 

\begin{figure}
\psfig{file=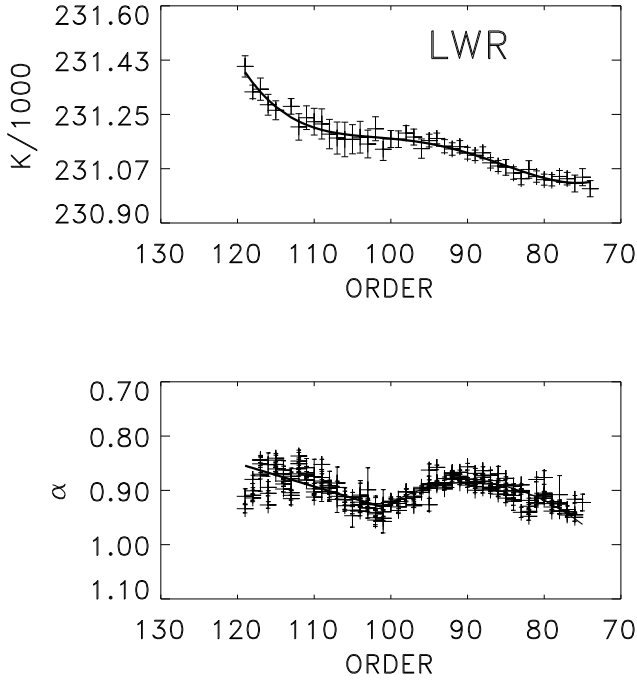,width=9cm}
\caption{The K and $\alpha$  parameters for the LWR camera as a
function of order number. The least-square polynomial fit to the
data as from Eq. \ref{eq:lwrkap} and Eq. \ref{eq:lwralp},
respectively, are also shown.}
\label{fig:lwrfig1}
\end{figure}

\begin{figure}
\psfig{file=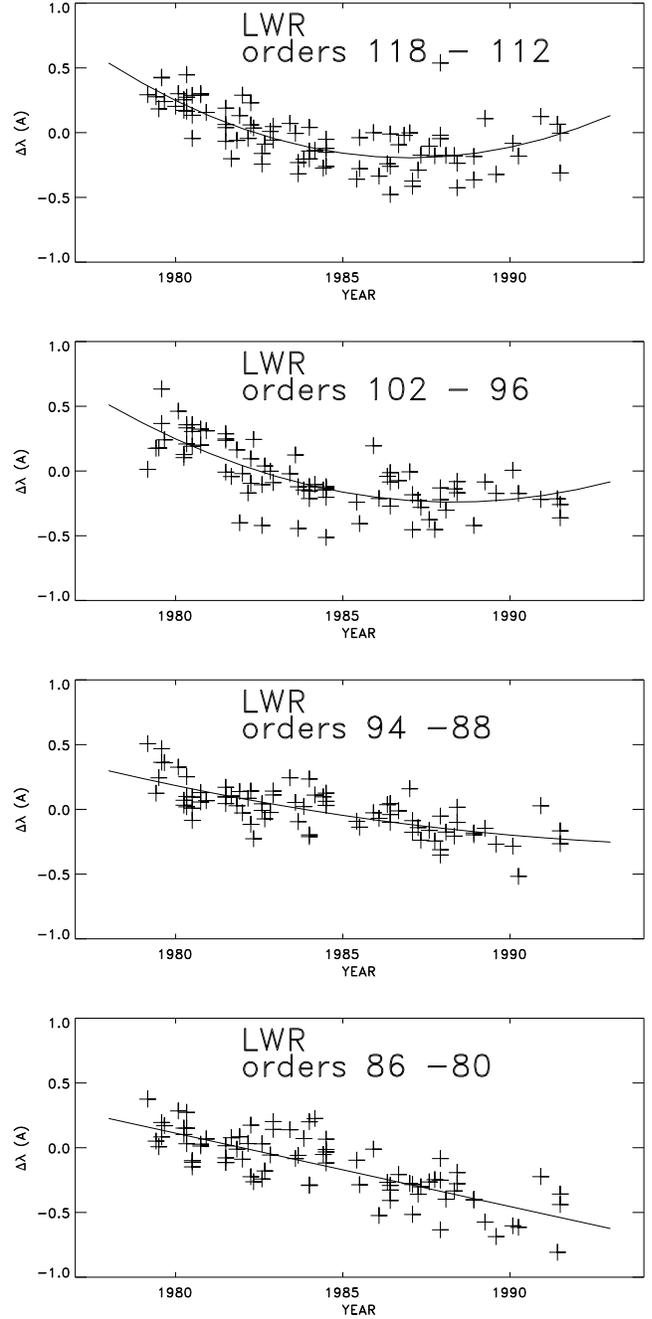,width=9cm}
\caption{Examples of how the wavelength shift 
$\Delta\lambda$=$\lambda_c-\lambda_{ref}$  for the LWR camera
depends on  the time of observations for four different ranges
of orders. The coefficients of the least-square polynomial fit
to the data are given in Table  \ref{tab:lwrtab}.}
\label{fig:lwrfig2}
\end{figure}

\paragraph{Dependence of $\lambda_{\rm c}$ on time:}
The central wavelengths obtained through Eq. \ref{eq:lwrkap} are
strictly valid only for the T=1982.6 reference date.  To study
the dependence of the central wavelengths on time we have used
86 LWR spectra of \iue\  calibration standards obtained between
1978 and 1994, and namely: 17 spectra of HD3360, 12 spectra of
HD34816 and 19 spectra of HD120315. The results indicate that,
differently from the LWP camera, the dependence on observing
time is generally non--linear, and it  varies across the camera
faceplate. We have computed the mean value
$\Delta\lambda$(m)=$\lambda_{\rm c}$(m)-$\lambda_{\rm ref}$(m)
in 10 overlapping windows, each containing seven consecutive
orders centered at

\begin{equation}
{\rm m_i}= 115-4({\rm i}-1) {\rm ~~~~~with~~i=1,10}
\label{eq:lwremme}
\end{equation}

\noindent
The averages were made from order m$_{\rm i}$-3 to order m$_{\rm
i}$+3. A quadratic fit to the data is appropriate in all cases
except near the long wavelength end of the camera, where a
linear behaviour is found. In summary, the wavelength shift
$\Delta\lambda=\lambda_{rm c}-\lambda_{\rm ref}$ can be
represented as:

\begin{equation}
\Delta\lambda{\rm(m_i,T)}={\rm a(m_i)+b(m_i)T+c(m_i)T^2}
\label{eq:lwrdel}
\end{equation}

\noindent
where T is the date of observation (fractional year). The
coefficients  applicable to individual orders are  obtained by
linear interpolation  of the data in Table \ref{tab:lwrtab}.  
For orders greater than  115 and smaller than 79 the
coefficients for m = 115 and m = 79 should be used,
respectively.

\begin{table}
\caption{Coefficients to determine the wavelength shifts for the 
LWR camera}
\begin{tabular}{c l l l}
\hline
Order   &       a(m$_i$)        &       b(m$_i$)        &       c(m$_i$)        \\
\hline
115     & 35736.699     &       -35.970790      &       0.0090515542    \\
111     & 62433.199     &       -62.859885      &       0.0158223297    \\
107     & 53287.133     &       -53.638642      &       0.0134980459    \\
103     & 42742.709     &       -43.014583      &       0.0108219635    \\     99      & 28040.843     &       -28.206111      &       0.0070930274    \\
95      & 10463.439     &       -10.501169      &       0.0026347077    \\
91      & 6223.1919     &       -6.2320255      &       0.0015601476    \\
87      & 4478.2512     &       -4.4662128      &       0.0011134034    \\
83      & 112.32014     &       -0.00566704     &       0               \\
79      & 156.16074     &       -0.0787367      &       0               \\
\hline
\end{tabular}
\label{tab:lwrtab}
\end{table}

\paragraph{Dependence of $\alpha$ on order number and focussing conditions:}
As in the case of the LWP camera, we find that  the ripple
parameter $\alpha$ has a strong dependence on the order number.
A polynomial fit to the data provides:

\begin{eqnarray}
\label{eq:lwralp}
\alpha & = & 1.360633-4.252626~10^{-3}{\rm m}  \\
& & {\rm (m=119-101)}  \nonumber \\
\alpha & = & 3.757863-0.0640201~{\rm m}+3.5664390~10^{-4}{\rm m}^2 \nonumber \\
& & {\rm (m=100-74)} \nonumber 
\end{eqnarray}

\noindent
with  rms errors of 0.02 and 0.01, respectively.

No significant correlation was found between the $\alpha$
parameter and the focussing conditions.

In Fig.~\ref{fig:lwrfig1} we show  K and $\alpha$ as a function
of order number for the mean date 1991.2, and in
Fig.~\ref{fig:lwrfig2} the mean wavelength shift $\Delta\lambda$
= $\lambda_{\rm c} -\lambda_{\rm ref}$ as a function of
observing date.

\begin{figure}
\psfig{file=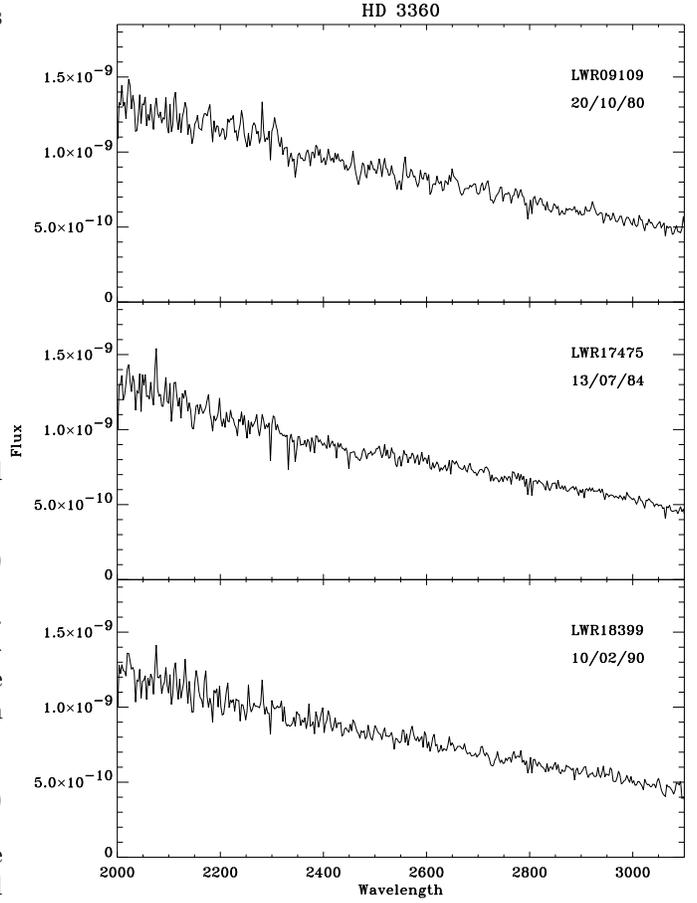,width=9cm}
\caption{Flux distribution of HD 3360 obtained from three LWR
ripple corrected and flux calibrated high resolution  rebinned
spectra obtained several years apart. Fluxes are in the same
units as in Fig. \ref{fig:swpridem}.}
\label{fig:lwrridem}
\end{figure}

It is interesting to note that, in spite of the strong time
dependence of the central positions of spectral orders (see Eq.
\ref{eq:lwrdel}), the  LWR ripple correction algorithm provides 
good  results even if applied to spectra obtained several years
apart, as shown in Fig. \ref{fig:lwrridem}.

\subsubsection {The ripple correction algorithm} 

In conclusion,  the blaze function for \iue\ high resolution
spectra processed with \newsips\ remains analytically defined by
Eq. \ref{eq:rip},  where the $\alpha$ parameter is  given in
Eqs. \ref{eq:swpalp}, \ref{eq:lwpalp} and \ref{eq:lwralp} for
SWP, LWP and LWR, respectively, and the central wavelengths  are
computed according to: \\

SWP: 
\begin{equation}
\lambda_{\rm c}(m)=137508.316/{\rm m}+0.0321729~{\rm THDA}+2.111841
\end{equation}
\noindent
which is the result of combining Eqs. \ref{eq:swpkap} and 
\ref{eq:swpdel}. \\

LWP: 

\begin{equation}
\lambda_{\rm c}(m)=230868.1770/\rm m-0.0263910~{\rm T}+56.433405
\end{equation}
which is the result of combining Eqs. \ref{eq:lwpkap} and 
\ref{eq:lwpdel}.\\

LWR: 

\begin{equation}
\lambda_{\rm c}(m)={\rm K(m)/m}+\Delta\lambda(\rm{m,T})
\end{equation}
where K is given by Eq. \ref{eq:lwrkap} and $\Delta\lambda$(m,T) is
computed according to Eq. \ref{eq:lwrdel}.

The ripple  correction algorithms were derived using spectral
orders 119 to 72 for the SWP camera, 118 to  74 for the LWP
camera and 118 to 76 for the  LWR camera. Extrapolation of the
algorithms to higher  and lower orders provides in general
satisfactory results.

\section{The absolute calibration of NEWSIPS high resolution spectra}
\label{sec:abscal}

For the absolute flux calibration of \newsips\ high resolution
spectra we have followed the method  described in Cassatella et
al. (1994).  Let us indicate with N($\lambda$) the
ripple-corrected high resolution Flux Numbers, normalized to the
exposure time. The corresponding absolute flux can be determined
from

\begin{equation}
{\rm F}(\lambda)=\rm C(\lambda)~\rm S^{-1}(\lambda)~\rm N(\lambda)~~{\rm erg/cm^2~s~\AA}
\end{equation}

\noindent
where S$^{-1}(\lambda)$ is the low resolution inverse
sensitivity function appropriate to the camera considered, and
C($\lambda$) is the so--called high resolution calibration
function defined as

\begin{equation}
\rm C(\lambda)=\rm n(\lambda)/\rm N(\lambda)
\end{equation}

\noindent
being n($\lambda$) the net Flux Numbers (i.e., not absolutely
calibrated), normalized to the exposure time, derived from low
resolution observations of the same target. Because of the
time--dependent sensitivity degradation of the cameras, the
pairs of low--high resolution spectra used to determine
C($\lambda$) should be obtained close enough in time.
Alternatively, both n($\lambda$) and N($\lambda$) should be
previously corrected for sensitivity degradation. This latter
approach, here followed, has the advantage of  increasing
considerably the number of usable spectra. The correction for
sensitivity degradation of high resolution spectra has been made
with  the same algorithms used for low resolution spectra, as
described by Garhart (1992, 1993) and Garhart et al. (1997).
This procedure is justified  in Paper III, which shows that high
resolution spectra obtained even several years apart, once
corrected in this way, provide very nearly the same flux
repeatability performance as spectra obtained close in time.

The high resolution spectra are first corrected for the blaze
function, resampled in 2 \AA\ bins, and normalized to the
exposure time in seconds.  The spectra are then corrected for
sensitivity degradation, for the THDA induced sensitivity
variations, and for the camera rise time following the same
algorithms used in the \newsips\ processing of low resolution
spectra (see Gonz\'alez-Riestra et al. 1999b, Paper IV).
Finally, the spectra of each target are averaged together to
obtain a mean spectrum N($\lambda$). To obtain C($\lambda$), the
mean high resolution spectra of a given target are then divided
by the mean low resolution spectrum of the same target (which
are also corrected for temperature effects, camera rise-time and
sensitivity degradation).

\subsection{SWP}

To determine the high resolution calibration function
C($\lambda$) we have used 28 SWP high resolution spectra of
BD+28 4211, 38 of BD+75 325, 27 of HD60753, 6 of G191 B2B and 13
spectra of CD-38 10980.  We find that the repeatability errors
on N($\lambda$) (after all the above mentioned corrections are
applied) are typically 3-5\%. These small errors confirm the
validity of applying the low dispersion degradation rates to
high dispersion spectra.

\begin{figure}
\psfig{file=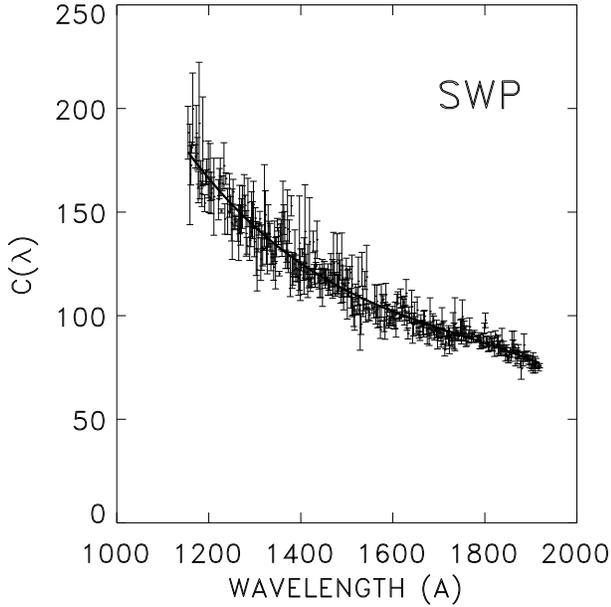,width=9cm}
\caption{The high resolution calibration function C($\lambda$)
for SWP spectra. Error bars are indicated. The thick line
represents the polynomial representation in Eq.
\ref{eq:clamswp}} 
\label{fig:clamswp} 
\end{figure}

The low resolution net fluxes n($\lambda$) of the above targets
were obtained by averaging many low resolution spectra obtained
during the 1990-1991 re--calibration period.  The curve
C($\lambda$) for the SWP camera is shown in Fig.
\ref{fig:clamswp}.

A third order polynomial fit to the data provides:

\begin{equation}
\rm C(\lambda)=\rm A+\rm B~\lambda+\rm C~\lambda^2+\rm D~\lambda^3
\label{eq:clamswp}
\end{equation}

\noindent where \\

\noindent 
A=1349.8538 \\
B=-2.0078566 \\
C=1.10252585~10$^{-3}$ \\
D=-2.0939327~10$^{-7}$ \\
\noindent
with an standard deviation of 6.3.  The above equation has been
derived  from data in the wavelength range 1175 to 1950 \AA. 

The repeatability error on C($\lambda$) is 4\%, irrespective of
wavelength, which we take as the internal error of the high
resolution calibration function.

\subsection{LWP}

\begin{figure}
\psfig{file=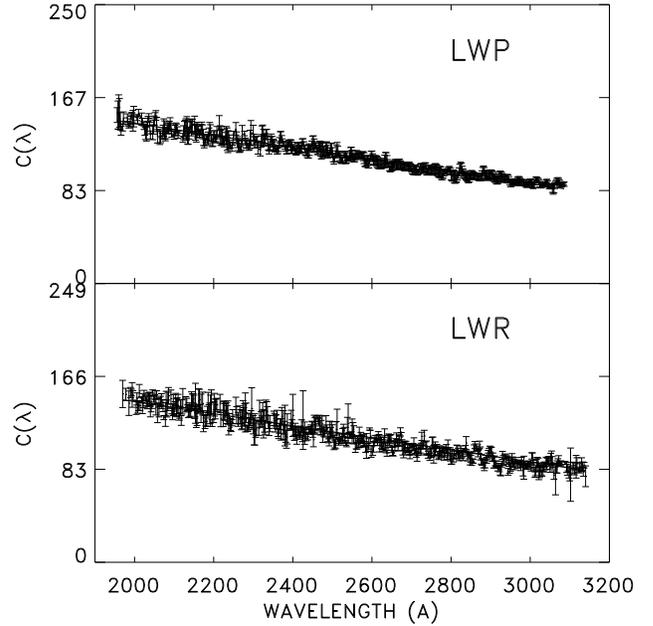,width=9cm}
\caption{The high resolution calibration function C($\lambda$)
for LWP  and LWR  spectra. Error bars are indicated. The thick line
represents the adopted linear representation of C($\lambda$) in
Eq. \ref{eq:clamlw}, which is the same for the LWP and LWR cameras.}
\label{fig:clamlw}
\end{figure}

To determine C($\lambda$) for the LWP camera we have used 25
high resolution spectra of BD+28 4211, 37 of BD+75 325, 27 of
HD60753 and 4 spectra of CD-38 10980.  The repeatability errors
on N($\lambda$) reach the 4\%\ level at 2400 \AA, but do not
exceed 2-3\% around 2800 \AA. Similarly to the case of the SWP
camera, these small errors confirm the applicability of the low
dispersion sensitivity degradation algorithm to high resolution
data.  The low resolution net fluxes n($\lambda$) of the above
targets were obtained by averaging many low resolution spectra
obtained during the 1990-1991 re--calibration period, extracted
with \newsips\ and corrected for time-dependent sensitivity
degradation according to Garhart (1993). The curve C($\lambda$)
for the LWP camera is shown in Fig. \ref{fig:clamlw}.  A linear
fit to the measurements provides:

\begin{equation}
\rm C(\lambda)=251.383956-0.053935103~\lambda
\label{eq:clamlw}
\end{equation}
\noindent
with a standard deviation of 3.49. The repeatability of the
C($\lambda$) function is about 4\%. The above equation has been
derived  from data in the wavelength range 1975 to 3150 \AA. 

\subsection{LWR}

To determine C($\lambda$) we have used a total of 17 high and 23
low resolution spectra of the calibration standards BD+28 4211,
BD+75 325, HD60753 and HD93521. 

We find that the repeatability errors on the net fluxes
N($\lambda$) after the sensitivity degradation correction are of
the same order as for the LWP camera, confirming once more the
applicability of the low dispersion sensitivity degradation
algorithm to high resolution spectra.

The resulting determinations of C($\lambda$) are reported in the
bottom panel of Fig. \ref{fig:clamlw}. It is interesting to note
that the data points are well fitted by the same analytical
representation as for the LWP camera (Eq. \ref{eq:clamlw} ). 
The residuals correspond to an rms error on C($\lambda$) of 5.3.
A linear fit would provide about the same residuals (4.5), and
the curve would only deviate from that of the LWP by 1.3\%. The
internal accuracy of the calibration function ranges from 5\%\
below 2300 \AA\ to 3\%\ at longer wavelengths. The wavelength
range covered by the LWR high resolution  absolute calibration
is the same as for the LWP camera.

Note that also Cassatella et al. (1994) found that the
C($\lambda$) curve is the same for LWP and LWR data processed
with \iuesips.

\subsection{Examples of application}

\begin{figure} 
\psfig{file=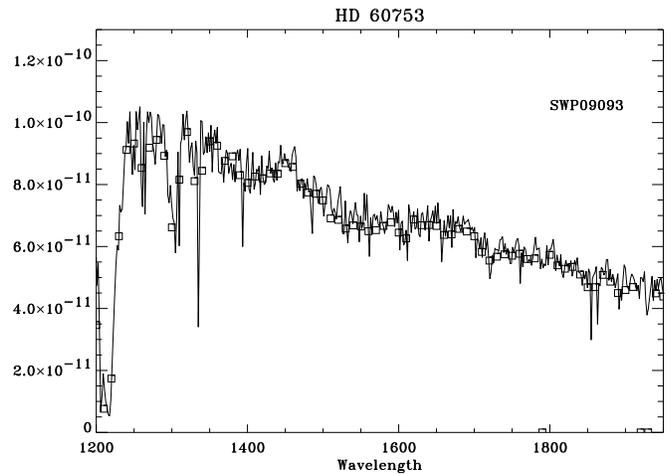,width=9cm}
\caption{The flux distribution of the \iue\ standard HD60753
obtained from the  high resolution spectrum SWP 9093 (full line)
is compared  with the absolute fluxes of the same star (open
boxes). The high resolution spectrum has been rebinned to the
low resolution wavelength step. Fluxes are in the same units as
in Fig. \ref{fig:swpridem}.} 
\label{fig:swpabs} 
\end{figure}

\begin{figure} 
\psfig{file=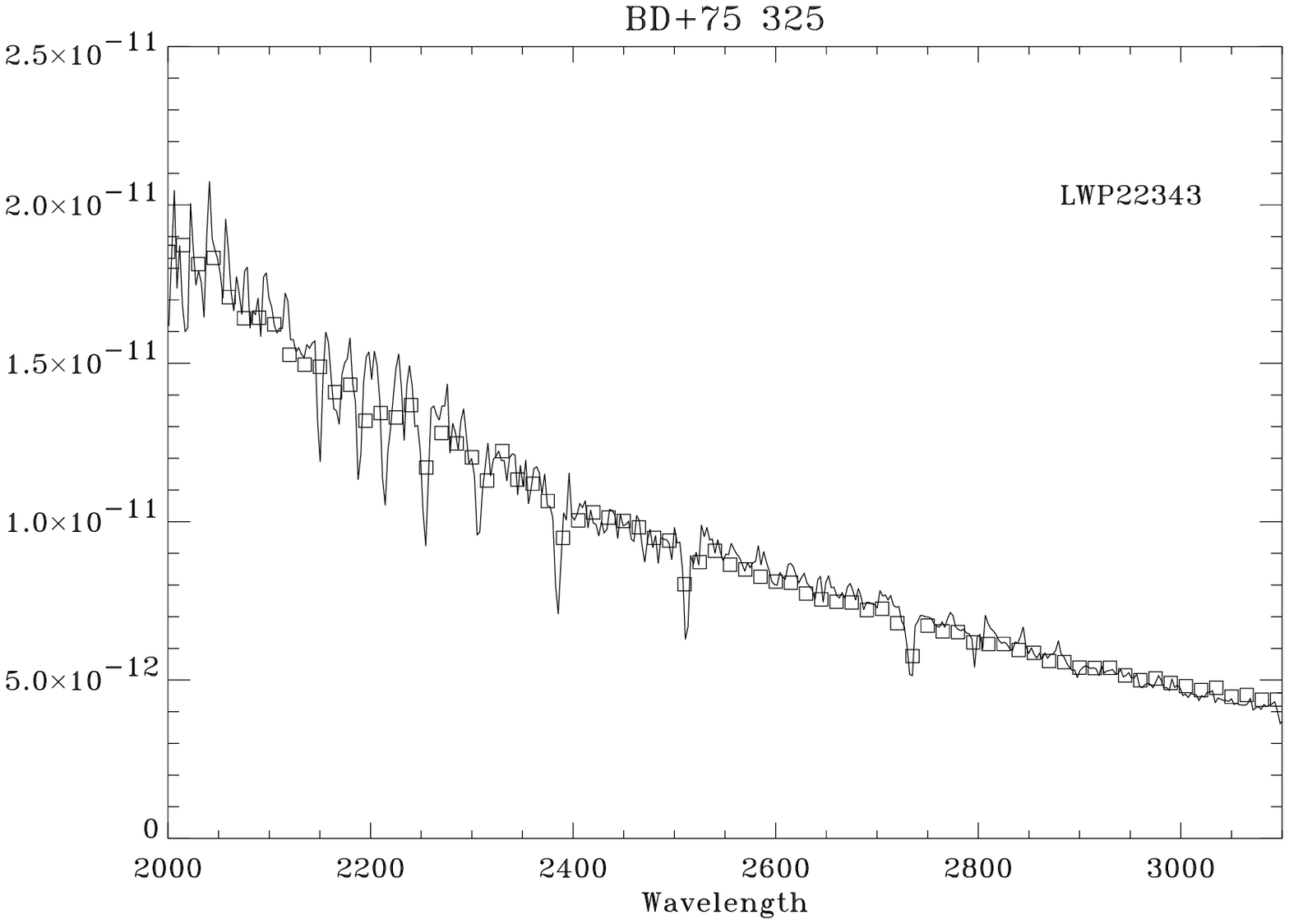,width=9cm} 
\caption{The flux distribution of the \iue\ standard BD+75 325
obtained from the high resolution spectrum LWP22343 (full line)
is compared  with the absolute fluxes of the same star (open
boxes). The high resolution spectrum as been rebinned to the low
resolution wavelength step. Fluxes are in the same units as in
Fig. \ref{fig:swpridem}.} 
\label{fig:lwpabs} 
\end{figure}

\begin{figure}
\psfig{file=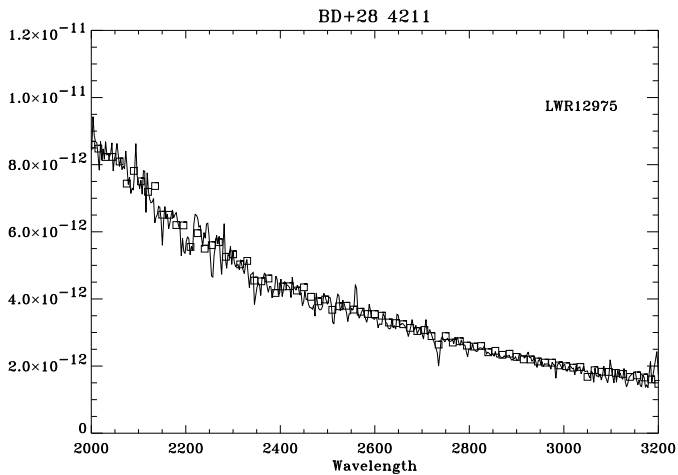,width=9cm}
\caption{The flux distribution of the \iue\ standard BD+28 4211
obtained from the  high resolution spectrum LWR12975 (full line)
is compared  with the absolute fluxes of the same star (open
boxes). The high resolution spectrum has been rebinned to the
low resolution wavelength step. Fluxes are in the same units as
in Fig. \ref{fig:swpridem}.}
\label{fig:lwrabs}
\end{figure}

It is important to compare the fluxes obtained from high
resolution spectra using the present method with  the fluxes of
the calibration standards which define the \iue\ flux scale
(Paper IV). Examples of such a comparison are given in Figs.
\ref{fig:swpabs}, \ref{fig:lwpabs} and \ref{fig:lwrabs}
referring to the stars HD60753, BD+28 4211 and BD+75 325
observed with the SWP, LWP and LWR cameras, respectively.

In the above examples, the agreement between high and low
resolution fluxes is within the 4\%\ repeatability errors quoted
above.

\begin{figure} 
\psfig{file=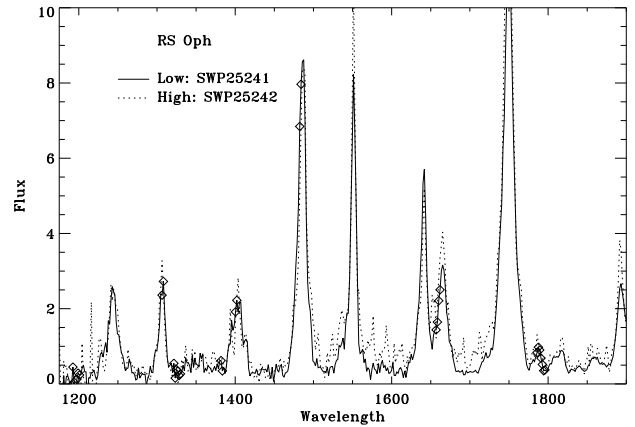,width=9cm}
\caption{Comparison between the flux calibrated high and low
resolution spectra of the recurrent nova RS Oph. Diamonds mark
flagged pixels (saturated or affected by reseau marks) in the
low resolution spectrum. Fluxes are in units of 10$^{-12}$ erg
cm$^{-2}$ s$^{-1}$ \AA$^{-1}$.} 
\label{fig:rsoph} 
\end{figure}

It should be stressed that the present calibration is applicable
to both continuum and emission line sources. This is confirmed
by  line emission measurements in several  pairs of low-high
resolution spectra of emission line sources.  As an example, we
show in Fig. \ref{fig:rsoph} a low and a high resolution
spectrum of the recurrent nova RS Oph taken very close in time.

\begin{figure} 
\psfig{file=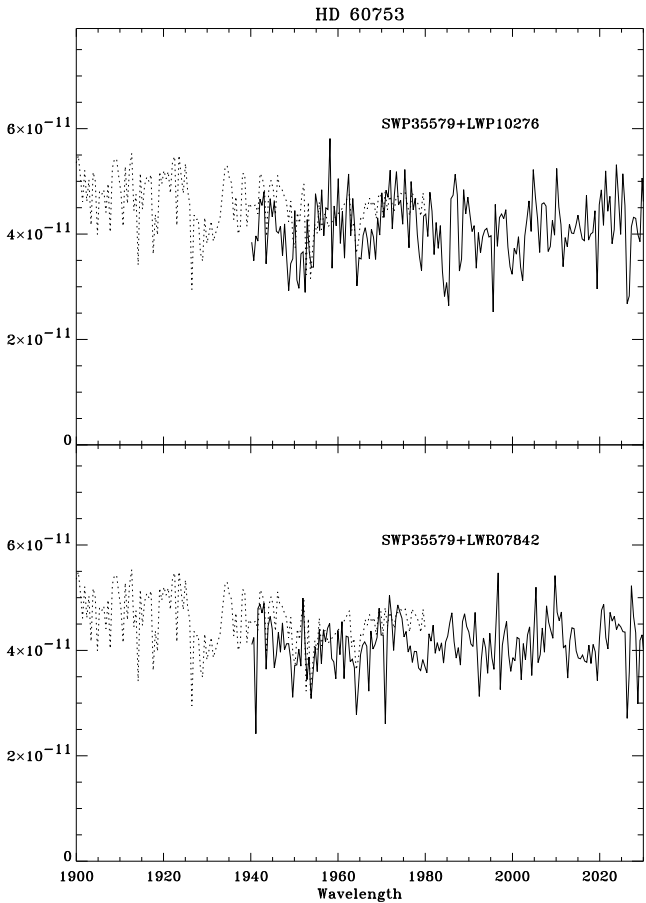,width=9cm} 
\caption{The combined SWP-LWP and SWP-LWR spectra of HD60753
are  shown in the overlap region around 1950 A. The original
data have been rebinned in 0.5 \AA\ steps. Fluxes are in the
same units as in Fig. \ref{fig:swpridem}.} 
\label{fig:over} 
\end{figure}

Another test made was to verify the accuracy of the absolute
calibration in the overlap region around 1950 \AA\ between the
SWP and the LWP and LWR cameras. We find that, in this region,
the short and long wavelength cameras agree to within 2 to 6\%\,
on average.  As an example, we show in Fig. \ref{fig:over} the
overlap region for two pairs of spectra of HD60753.

\begin{figure}
\psfig{file=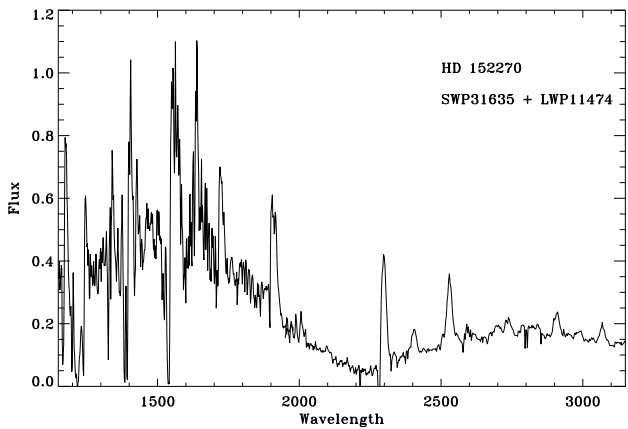,width=9cm}
\caption{Combined SWP--LWP spectrum of the Wolf Rayet star
HD152270 after application of the ripple correction and absolute
calibration algorithms here described. The high resolution
spectra have been resampled to the low resolution wavelength
step for display purposes. Fluxes are in units of 10$^{-12}$ erg
cm$^{-2}$ s$^{-1}$ \AA$^{-1}$.} 
\label{fig:wr} 
\end{figure}

The good match between short and long wavelength high resolution
spectra can also be deduced from  Fig. \ref{fig:wr}, which shows
a combined SWP-LWP spectrum of the Wolf Rayet star HD152270.

We note that residual non-linearity effects in the Intensity
Transfer Function, especially in the case of underexposed
spectra, or spectra near the saturation limit, can occasionally
cause a flux mismatch between the short and long wavelength
cameras.

\section{Conclusions}
\label{sec:conclusi}

We have shown  that a reliable correction for the echelle blaze
function can be obtained for \iue\ high resolution spectra
processed with  \newsips. We find that not only the K
parameter,  but also the $\alpha$ parameter varies with order
number, and that the central peaks of the blaze function deviate
from the reference positions in a way which depends on the value
of the THDA temperature for the SWP camera, and on the
observation date for the LWP and LWR cameras. Having taken into 
account all these dependences (and not only the dependence of K
on order number as in previous investigations), allows the
present algorithm to successfully correct most spectra.
According to our experience, the algorithm might fail in a few
cases (probably less then 5-10\%). The identified cases of
failure correspond to images in which the interorder background
level is not properly evaluated with the \newsips\ automatic
extraction procedure or to cases in which the target is not
acquired (purposely or not) in the center of the large entrance
apertures of the \iue\ spectrographs. In this respect, we note
that a displacement of 1 arcsec along the dispersion line
corresponds to a velocity shift of the wavelength scale of about
5 km/s, irrespective of the camera used. For example, a
displacement of 5 arcsec, or  0.125 \AA\ at 1500 \AA\, would
produce a flux mismatch in the overlap region midway between
adjacent orders of about 4\%.

As for the absolute calibration of \iue\ high resolution
spectra, we have shown that its internal accuracy is  about 
4\%\,  and that it applies both to continuum and emission line
sources.  The flatter shape of the C($\lambda$) function found
here is due to the improved background extraction procedure in
\newsips\ (Smith 1999).

\begin{acknowledgements}

We would like to acknowledge the contribution of all VILSPA
staff to the development and production of the \ines\ system and
the  referee, Dr. Joy S. Nichols, for useful comments. 

\end{acknowledgements}

\end{document}